\documentclass[aps,pre,reprint,superscriptaddress]{revtex4-1}
\usepackage{graphicx,amssymb}
\usepackage{color,ulem}

\newcommand{\bk}{\mathbf{k}}
\newcommand{\bl}{\mathbf{l}}
\newcommand{\kb}{k_{\text{B}}}
\newcommand{\eqref}[1]{(\ref{#1})}

\renewcommand{\Im}{\mathop \mathrm{Im}}
\newcommand{\av}[1]{\langle #1 \rangle}
\newcommand{\hf}{\frac{_1}{^2}}
\newcommand{\caO}{{\mathcal O}}

\begin{document}

\title{Energy fluctuations of finite free-electron Fermi gas}


\author{Jukka P. Pekola}
\affiliation{Low Temperature Laboratory, Department of Applied Physics, Aalto University School of Science, P.O. Box 13500, 00076 Aalto, Finland}
\author{Paolo Muratore-Ginanneschi} 
\author{Antti Kupiainen}
\affiliation{University of Helsinki, Department of Mathematics and Statistics, P.O. Box 68 FIN-00014, Helsinki, Finland}
\author{Yuri M. Galperin}
\affiliation{Department of Physics, University of Oslo, PO Box 1048 Blindern, 0316 Oslo, Norway}
\affiliation{Ioffe Institute, 26 Politekhnicheskaya, St Petersburg 194021, Russian Federation}

\date{\today}

\begin{abstract}
We discuss the energy distribution of free-electron 
Fermi-gas, a problem with a textbook solution of Gaussian energy fluctuations in the limit of a large system. We find that for a small system, characterized solely by its heat capacity $C$, the distribution can be solved analytically, and it is both skewed and it vanishes at low energies, exhibiting a sharp drop to zero at the energy corresponding to the filled Fermi sea. The results are relevant from the experimental point of view, since the predicted non-Gaussian effects become pronounced when $C/k_B \lesssim 10^3$ ($k_B$ is the Boltzmann constant), a regime that can be easily achieved for instance in mesoscopic metallic conductors at sub-kelvin temperatures.
\end{abstract}


\maketitle

\section{Introduction}

Physical quantities of an equilibrium macroscopic system are well characterized by their average values. However, random deviations from the average values - fluctuations - are very important since they contain important information on the system. Under most circumstances the distribution of small fluctuations is Gaussian~\cite{LL}, ch.~7. 
However, this is not the case for small devices, which are currently studied intensively~\cite{Jarzynski1997,Crooks1999},
see~\cite{rmp06,Rowe,Campisi2011,Seifert2012} for a review. Energy and temperature fluctuations in a single electron box~\cite{IN} were considered 
in~\cite{Samuelsson}; several works experimental and theoretical, e.g.,~\cite{gasparinetti,Viisanen,Heikkila}, were devoted to temperature fluctuations.

In the present work we will derive the distribution function for a finite Fermi gas and show that its shape is determined by only one dimensionless parameter - the total heat capacity $C$ divided by the Boltzmann constant, $\kb$. At  $C/\kb \gg 1$ we recover the well-known Gaussian distribution, while at finite values of this parameter significant deviations are expected. We will analytically derive the distribution of the energies of a finite sample of the Fermi gas kept at a given temperature and analyze its properties including moments and skewness absent in the thermodynamic limit. Since the heat capacity of a metallic conductor of sub-micron dimensions at standard sub-kelvin experimental temperatures is of the order of $(10^2-10^3)\kb$ \cite{rmp06,jpnjp13}, our results have potential impact on sensitive bolometers \cite{richards,rmp06,gasparinetti}.   
In future, we plan to use the obtained distribution for analysis of heat exchange between a quantum device and a mesoscopic metallic calorimeter.

The paper is organized as follows. We derive general expression for the energy distribution in Sec.~\ref{general}, which will be analyzed in Sec.~\ref{different}. Moments of the distribution are considered in Sec.~\ref{moments}.

\section{Energy distributions}
\subsection{General expressions} \label{general}
It is convenient to calculate the Fourier transform, $F(\lambda)$, of the energy distribution function. It can be expressed in the form
\begin{equation} \label{cf}
F(\lambda) =\left \langle e^{i\lambda E} \right \rangle =\frac{\sum_m e^{(i\lambda -\beta) E_m} }{\sum_m e^{ -\beta E_m}}=\frac{Z(\beta - i \lambda)}{Z(\beta)}\, .
\end{equation}
Here 
\begin{equation}
Z(\beta)  \equiv  \sum_m e^{ -\beta E_m}
\end{equation}
is the partition function, $\beta \equiv (\kb T)^{-1}$ where $T$ is temperature,  while
\begin{equation} \label{en1}
E_m =\sum_\bk (\epsilon_{\bk,m} - \mu) \, n_{\bk,m} \, .
\end{equation}
A micro-state is characterized by the set of quantum numbers  $\{\bk,m\}$ where $\bk$ is the quasi-momentum. 
In the following, concentrating on mesoscopic electronic systems, we will consider a gas of elementary particles characterized only by the quantum number $\bk$. According to the definitions~\eqref{cf}-\eqref{en1}, we calculate the characteristic function of quasiparticle energy.

The energy distribution function, $P(E)$, can be calculated as the inverse Fourier transform of $F(\lambda)$ as
\begin{equation}
P(E)=\int_{-\infty}^\infty \frac{d\lambda}{2\pi} \, e^{-i\lambda E} F(\lambda)\, .
\end{equation}
For the Fermi gas~\cite{LL},
\begin{eqnarray}
\ln Z(\beta)&=& \sum_\bk\ln \left( 1+e^{-\beta(\epsilon_\bk - \mu)}\right) \\
&=&\mathcal{V} \int_0^\infty d\epsilon \, N(\epsilon) \, \ln \left( 1+e^{-\beta(\epsilon - \mu)}\right) \, .
\end{eqnarray}
Here $N(\epsilon)$ is the density of states.  In the following we put $N(\epsilon) = \text{const} \equiv N_0$. 
This assumption is valid for the two-dimensional gas, while in the three-dimensional case $N(\epsilon) \propto \sqrt{\epsilon}$. One can show that in  the limiting case $\beta \mu \gg 1$ the results differ only by renormalization of numerical constants in the expressions for the average energy and heat capacity.
For realistic experimental conditions $\beta \mu = T_F/T \sim 10^6$ for a metal with Fermi temperature $T_F \sim 10^5$ K at $T=0.1$ K. 

For $N(\epsilon)= N_0$ we obtain
\begin{eqnarray}
\ln Z(\beta)&=&\frac{\mathcal{V}N_0}{\beta} \int_{-\beta \mu}^\infty d\xi \ln \left  (1+e^{- \xi} \right ) \nonumber \\
&=&-\frac{\mathcal{V}N_0}{\beta}\, \mathrm{dilog}\, \left(1+e^{\beta \mu} \right)\, .
\end{eqnarray}
Here
\begin{equation}
\mathrm{dilog}\, (x) = \int_1^x \frac{\ln(t)}{1-t}\, dt 
\end{equation}
is the dilogarithm function. 
At $\beta \mu \gg 1$ we expand
\begin{equation} \label{exp}
- \mathrm{dilog}\, \left(1+e^{\beta \mu} \right) \approx \frac{1}{2} (\beta \mu)^2 + \frac{\pi^2}{6}-e^{-\beta \mu} + \ldots\, .
\end{equation}
Then
\begin{equation} \label{exp1}
\ln Z(\beta) \approx \mathcal{V}N_0 \left( \frac{1}{2}\beta \mu^2 + \frac{\pi^2}{6}\frac{1}{\beta}\right)\, .
\end{equation}
Note that we assume that the chemical potential is fixed rather than determined by the normalization of the single-particle distribution to the total number of particles. This assumption is appropriate for an electronic system with contacts.

In the case when $N(\epsilon) \propto \sqrt{\epsilon}$ we can put $N(\epsilon) = N_0\sqrt{\epsilon/\mu}$ to obtain instead of Eq.~\eqref{exp1}
\begin{equation}
\ln Z(\beta) \approx \mathcal{V}N_0 \left( \frac{2}{5}\beta \mu^2 + \frac{\pi^2}{4}\frac{1}{\beta}\right)\, .
\end{equation}
Therefore, different forms of the dependence of the density of states upon the energy 
indeed result only in numerical constants defining the average energy and heat capacity.
The difference in the numerical factors is compatible with the well known 
Sommerfeld expansion, see, e.g.,~\cite{FW}.

Using the expansion \eqref{exp1} we obtain 
\begin{eqnarray}\label{lnF}
\ln F(\lambda) &=&\ln \frac{Z(\beta - i\lambda)}{Z(\beta) } \nonumber \\
&=&\frac{ \mathcal{V}N_0}{2} \left(-i \lambda \mu^2+ \frac{\pi^2}{3} \frac{i\lambda}{\beta(\beta-i\lambda)}\right).
\end{eqnarray}
The first item is responsible for the energy shift of the energy by the Fermi energy $E_0 =\mathcal{V} N_0\mu^2/2$. In our approximation the specific heat has the standard free Fermi-gas expression 
(for a given chemical potential)
\begin{equation}
C=\kb \frac{\pi^2}{3\beta}N_0\mathcal{V}\, .
\end{equation}
Then the quantity in the parentheses in Eq.~\eqref{lnF} can be rewritten as
$$ i \lambda E_0 + \frac{C}{2\kb} \frac{i\lambda}{\beta - i\lambda}\, .$$
We rewrite 
\begin{eqnarray*}
\frac{C}{2\kb} \frac{i\lambda}{\beta -i\lambda}
&=&i\lambda \frac{C}{2\kb \beta}\left(\frac{i\lambda}{\beta -i\lambda} +1 \right) - i\lambda \frac{C}{2 \kb \beta}\\
&\equiv & -  i\lambda  E_\beta +\frac{C}{2\kb \beta} \frac{(i\lambda)^2}{\beta - i \lambda}. 
\end{eqnarray*}
Here $E_\beta = C/2 \kb\beta$. Denoting $\delta E =E-E_0- E_\beta$ we obtain in the exponent
\begin{equation}
-i\lambda \, \delta E -\frac{C}{2 \kb}\frac{\lambda^2}{\beta(\beta - i \lambda)}\, .
\end{equation}

Let us measure the energy deviation $\delta E$ in units of $\beta^{-1}\sqrt{C/\kb}$.  Then we put
\begin{equation}
\lambda \equiv  \frac{\beta}{\sqrt{C/\kb}} \Lambda , \ \delta E \equiv \frac{ \sqrt{C/\kb}}{\beta} u, \
c \equiv \frac{C}{\kb}
\end{equation}
to obtain for the second term
$$-\frac{1}{2}\frac{\Lambda^2}{1-i\Lambda/\sqrt{c}}= -\frac{1}{2}\frac{\Lambda^2(1+i\Lambda/\sqrt{c})}{1+\Lambda^2 /c}\, .$$
%
Therefore, we arrive at the expression for the distribution of dimensionless energies $u$:
\begin{equation} \label{Pu}
P(u)  =  \int_{-\infty}^\infty \frac{d\Lambda}{2\pi}\exp\left[ i\Lambda\left(\frac{1}{2}\frac{i\Lambda}{1-i\Lambda/\sqrt{c}}-u \right)\right]\, .
\end{equation}
This is our final expression, which we have to analyze for different values of dimensionless heat capacity, $c$.
\subsection{Energy distributions for different dimensionless heat capacity} \label{different}

We compute the integral  in Eq.~\eqref{Pu} by deforming the contour in the complex plane. The integrand is analytic in the full complex plane except for (an essential) singularity at $\Lambda=-i\sqrt{c}$. For large $|\Lambda|$  the argument of the exponential behaves as
$$
-\frac{1}{2}\frac{\Lambda^2}{1-i\Lambda/\sqrt{c}}-i\Lambda u =
-\frac{c}{2}-i\Lambda\left(\frac{\sqrt{c}}{2}+u\right)+\caO(1).
$$
 For thermodynamics it suffices to consider energies above the Fermi energy $E>E_0$ which means  $u+\sqrt{c}/2 >0$. We may then deform the contour to the lower half plane $\Im \Lambda<0$. We get
 $$
P(u)=P_{0}(u)+P_\infty(u)
$$
where $P_\infty(u)$ is contribution from the semi circle at infinity in the lower half plane and
$P_{0}(u)$ from a circle around $\Lambda=-i\sqrt{c}$. 

Writing 
 $\Lambda=\rho e^{i\phi}$,  $\pi \le \phi \le  2\pi$  we obtain $P_\infty(u)$ as the limit of
\begin{equation}
e^{-\hf c}\frac{i\rho}{2\pi}\int_{2\pi}^\pi d\phi e^{i(\phi -\rho b e^{i\phi})}=\frac{\sin(b\rho)}{\pi b}
\end{equation}
as $\rho\to\infty$ i.e. $P_\infty(u)$ where $ b=\sqrt{c}/2+u$.  Since we assume $b\neq 0$ this limit vanishes.

To compute $P_{0}(u)$ change variables to $z=\frac{1}{c}(1-i\frac{\Lambda}{\sqrt{c}})$ to get
$$
P_{-i}(u) =c^{\frac{3}{2}}e^{-(2c+\sqrt{c}u)}\oint e^{\frac{1}{z}}e^{xz}\frac{dz}{2\pi i}
$$
where $x=c(c+\sqrt{c}u)$ and the integral is around the origin. Then 
\begin{eqnarray}
&&\oint e^{\frac{1}{z}}e^{x z}\frac{dz}{2\pi i}=\sum_{n=0}^\infty \frac{1}{n!} \oint z^{-n}e^{x z}\frac{dz}{2\pi i}\nonumber\\&=&\sum_{m=0}^\infty \frac{1}{(m+1)!}\frac{1}{m!}x^{m}=x^{-\hf}I_1(2x^\hf)\nonumber
\end{eqnarray}
where $I_1$ is the modified Bessel function.

As a result,
\begin{eqnarray}
P (u)& =&\frac{c^{3/4}}{\sqrt{\sqrt{c}+2u}} I_1 \left(c^{3/4}\sqrt {\sqrt{c}+2u}\right)e^{-u\sqrt{c} -c} 
\nonumber
\end{eqnarray}

\begin{figure}[t]
\centering
\includegraphics [width=\columnwidth] {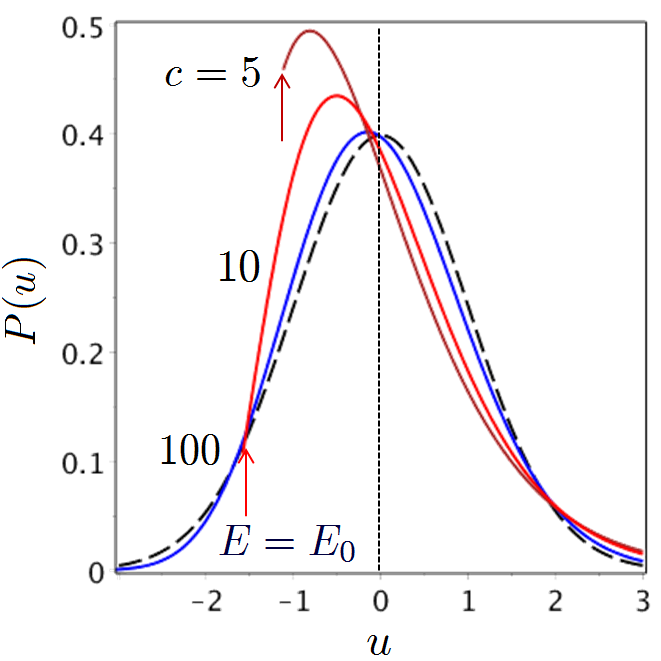}
\caption{(Color online) Distribution $P(u)$, where $u= \beta \delta E/\sqrt{c}$, where $\delta E$ 
is the deviation of the energy of the Fermi-gas from its mean, for a few different values of $c = C/k_B$ (solid lines).  
Dashed line is the Gaussian distribution.
\label{fig2}}
\end{figure}

Graphs for $P(u)$ for different heat capacities, $c$, are shown in Fig.~\ref{fig2}. 
One observes that at large $c$ the distribution is essentially Gaussian, while at small $c$ it becomes very asymmetric.

Let us crudely estimate the dimensionless heat capacity, $c$. The estimate reads as
$c \sim n_e \mathcal{V} (\kb T/\mu)$ where $n_e$ is the electron density. For a typical metal $n_e \approx 10^{23} \ \text{cm}^{-3}$. 
Assuming the sizes of a mesoscopic system (in $\mu\text{m}$) as 
$(0.3-1)\times (0.05-0.1)\times (0.01-0.03)$ $(\mu\text{m})^3$, we get 
$$\mathcal{V} = (1 .5\cdot 10^{-16} - 3\cdot 10^{-15}) \ \text{cm}^3\, . $$
and the total number of electrons is 
$$\mathcal{N} \equiv n_e\mathcal{V} =1.5\cdot 10^7 - 3\cdot 10^8\, .$$
Putting $T_{\text{F}} \approx 5\cdot10^4~\text{K}$ and $T=(0.05 - 0.1)~\text{K}$ we get
$$\kb T/\mu =T/T_{\text{F}}=(1-2)\cdot 10^{-6}.$$
Consequently, the range of the quantity $c$ is $15 -600$. As we have seen, at these values of $c$ the energy distribution is clearly non-Gaussian.

\subsection{Moments of the energy distribution} \label{moments}

The moments of the energy distribution can be readily calculated either from the Eq.~\eqref{Pu}, 
or explicitly from the Hamiltonian$ \mathcal{H}=\sum_\bk \epsilon_\bk a^\dag_\bk a_\bk$. In particular,
\begin{equation}
\langle E \rangle =\sum_\bk \epsilon_\bk \av{ a^\dag_\bk a_\bk} =\sum_\bk \epsilon_\bk f(\epsilon_\bk)
\end{equation}
where $f(\epsilon)=\left[e^{\beta(\epsilon - \mu)}+1 \right]^{-1}$ is the Fermi function. The second moment reads
\begin{equation}
\av{E^2}=\sum_\bk \epsilon_\bk \epsilon_\bl \av{a^\dag_\bk a_\bk a^\dag_\bl a_\bl}
=\av{E}^2 + \sum_\bk \epsilon_\bk^2 f( \epsilon_\bk)[1-f( \epsilon_\bk)].
\end{equation}
Here we have decomposed the product of four Fermi operators according to the Wick theorem.

We observe that only the vicinity of the Fermi level is important for the difference 
$$\av{\delta E^2} \equiv \left\langle \left(E-\av{E}\right)^2\right \rangle =\av{E^2}-\av{E}^2\, . $$
Therefore, while calculating $\av{\delta E^2} $ we can assume constant density of states and $\beta \mu \gg 1$. 
In this way we obtain
\begin{equation}
\av{\delta E^2} =\kb T^2 C\, .
\end{equation}
In a similar way, we calculate
\begin{equation}
\av{\delta E^3}=2\sum_\bk \epsilon_\bk^3 f( \epsilon_\bk)[1-f( \epsilon_\bk)]^2=3\kb^2T^3C\, .
\end{equation}
The skewness of the distribution is then
\begin{equation}
\gamma =\av{\delta E^3}/\av{\delta E^2}^{3/2} =3/\sqrt{c}\, .
\end{equation}
Obviously, the skewness vanishes as $c \to \infty$ which is the thermodynamic limit. 
For a Gaussian distribution, the relationship between the 2nd and the 4th moments is 
$\av{\delta E^4}=3\av{\delta E^2}^2$. In the general case, an additional contribution appears, such that
\begin{equation}
\av{\delta E^4}=
3\av{\delta E^2}^2\left[1+\frac{4}{\sqrt{c}} \right]\, .
\end{equation}
The 5th moment, absent in the Gaussian approximation, is
\begin{equation}
\av{\delta E^5} = 60\kb^4T^5 C\, .
\end{equation}
The results obtained by numerical integration of Eq.~\eqref{Pu} agree with the analytic expressions given above.

In summary, we have shown that the energy distribution of a free Fermi-gas with small heat capacity 
is non-Gaussian, with a sharp cut-off at low energies. This is a natural consequence of the minimal energy 
of the filled Fermi sea. We find that the heat capacities demonstrating strong non-Gaussian features can 
be achieved in standard metallic nanodevices at sub-kelvin temperatures.


\begin{acknowledgments}
We thank Ivan Khaymovich and Kay Schwieger for discussions. YMG thanks Aalto University for hospitality during the preparation of this manuscript. 
The work has been supported by the Academy of Finland (contracts no. 272218, 284594 and 271983). 
\end{acknowledgments}


\end{document}